\documentclass[comment, 
twocolumn,nofootinbib, preprintnumbers, superscriptaddress]{revtex4}

\usepackage{amsmath,amssymb,bm,slashed,braket}
\usepackage{graphicx}
\usepackage{epstopdf}
\usepackage{float}
\usepackage[colorlinks=true,
            linkcolor=blue,
            urlcolor=blue,
            citecolor=green,          
            bookmarks=true,
            bookmarksnumbered=true,
            breaklinks=true,
            pdfpagemode=Fullscreen,
            pdfstartview=FitBH]{hyperref}

\usepackage[normalem]{ulem}
\usepackage{color}

\definecolor{Orange}{cmyk}{0,0.61,0.87,0}
\definecolor{JungleGreen}{cmyk}{0.99,0,0.52,0}
\definecolor{OliveGreen}{cmyk}{0.64,0,0.95,0.40}
\definecolor{Brown}{cmyk}{0,0.81,1,0.60}
\definecolor{RoyalBlue}{cmyk}{0.71,0.53,0,0.12}
\definecolor{Gray}{cmyk}{0,0,0,0.40}
\definecolor{LightPink}{cmyk}{0.0,0.25,0,0}
\definecolor{LLightPink}{cmyk}{0.0,0.10,0,0}
\definecolor{LightBlue}{cmyk}{0.25,0,0,0}
\definecolor{LightGray}{cmyk}{0,0,0,0.2}

\usepackage{xcolor}
\definecolor{gesfpurple}{rgb}{0.47,0.19,0.42}

\definecolor{gesflanse}{rgb}{0.00,0.50,0.50}

\definecolor{gesfblue}{rgb}{0.08,0.42,0.76}

\definecolor{gesfred}{rgb}{1,0,0}

\definecolor{gesfwhite}{rgb}{1,1,1}

\definecolor{gesfblack}{rgb}{0,0,0}

\newcommand{\geqn}[1]{Eq.\,\hypersetup{linkcolor=blue}(\ref{#1})\hypersetup{linkcolor=blue}}
\newcommand{\gfig}[1]{{\hypersetup{linkcolor=violet}Fig.\,\ref{#1}\hypersetup{linkcolor=blue}}}

\graphicspath{{figs/}}

\begin{document}

\title{
Incoherent fermionic dark matter absorption with nucleon Fermi motion
}

\author{Shao-Feng Ge}
\email{gesf@sjtu.edu.cn}
\affiliation{Tsung-Dao Lee Institute \& School of Physics and Astronomy, Shanghai Jiao Tong University, China}
\affiliation{Key Laboratory for Particle Astrophysics and Cosmology (MOE) \& Shanghai Key Laboratory for Particle Physics and Cosmology, Shanghai Jiao Tong University, Shanghai 200240, China}

\author{Oleg Titov}
\email{titov\_o@sjtu.edu.cn}
\affiliation{Tsung-Dao Lee Institute \& School of Physics and Astronomy, Shanghai Jiao Tong University, China}
\affiliation{Key Laboratory for Particle Astrophysics and Cosmology (MOE) \& Shanghai Key Laboratory for Particle Physics and Cosmology, Shanghai Jiao Tong University, Shanghai 200240, China}

\begin{abstract}
We investigate the incoherent regime of the fermionic dark matter
absorption by nuclei using the relativistic Fermi gas model and
nuclear
form factors.
With the momentum transfer being roughly equal to the dark matter mass $m_\chi$, the incoherent regime contributes significantly to the absorption process for $m_\chi \gtrsim 100$\,MeV
with a spin-independent operator and for even smaller mass with
a spin-dependent one.
We also compare the situations for
various target nuclei
($^{131}$Xe, $^{72}$Ge, $^{40}$Ar,
$^{20}$Ne and $^4$He) that are typically used in the
dark matter direct detection. A heavier nucleus
actually has the
advantage of probing the incoherent scattering of the fermionic
absorption dark matter. Observing both the coherent and
incoherent contributions would be an important justification
of the fermionic dark matter absorption.
\end{abstract}

\maketitle

\section{Introduction}
More than 80\% of the matter in our Universe today 
is dark matter (DM) \cite{Arbey:2021gdg,Young:2016ala}.
But its nature still remains unknown. The direct detection
\cite{Cooley:2021rws,Misiaszek:2023sxe}
experiments searching for the DM particle candidates have already
explored a substantial region of the parameter space in the
1\,GeV $\sim$ 1\,TeV mass range
\cite{Billard:2021uyg,Aalbers:2022dzr}.
For sub-GeV DM, the energy deposit by
the DM scattering off a nucleus
is typically below the experimental threshold
\cite{Kahn:2021ttr,Essig:2022dfa}.
This has motivated a new wave of DM model building
studies to seek viable mechanisms
that could provide a detectable signal.

From this point of view, the DM absorption scenarios
where the DM mass is efficiently converted to recoil
energies are particularly attractive for probing
sub-GeV DM. Both bosonic \cite{Pospelov:2008jk, An:2014twa, Bloch:2016sjj,Hochberg:2016ajh,
Hochberg:2016sqx,Green:2017ybv,Arvanitaki:2017nhi,vonKrosigk:2020udi,
Mitridate:2021ctr,Hochberg:2021zrf} and fermionic
\cite{Dror+_2019a, Dror+_2019b, Dror+_2020, Ge+_2022,Ge:2023wye}
DM can be absorbed. The bosonic DM can be fully
absorbed in processes such as the dark photoelectric
effect $a + e \rightarrow e$
to ionize the atomic electron and the inverse Primakoff
process $a + {}^A_Z X \rightarrow \gamma + {}^A_Z X$
or the Compton-like
scattering $a + e \rightarrow \gamma + e$ that converts
an axion ($a$) to a photon. Since both electron and photon
are visible, all the axion energy can deposit in detector.

In comparison, the fermionic DM absorption requires
a fermion in the final state to conserve the fermion
number or equivalently the angular momentum.
This does not necessarily mean that
only part of the fermion mass can convert to deposit
energy. To be more precise,
the final-state particles
of the charged current processes
$\chi + {}^A_Z X \rightarrow e^\pm + {}^A_{Z \mp 1}X^*$
are all visible in the DM direct detection detector
\cite{Dror+_2019a, Dror+_2019b}.
On the other hand, the neutral current version
$\chi + {}^A_Z X \rightarrow \nu + {}^A_Z X$ has
a neutrino ($\nu$) in the final state and hence can deposit
only a fraction of the DM mass as target recoil energy
\cite{Dror+_2019a, Dror+_2019b, Dror+_2020, Ge+_2022, Li:2022kca}.
Nevertheless, the mass to deposit energy
conversion rate is still much higher than the classic
elastic scattering. Both nuclei
\cite{Gu+_2022,Arnquist+_2022,Dai+_2022}
and electron \cite{Zhang+_2022,Al_Kharusi+_2023,Liu+_2024} targets
have been searched in direct detection experiments.
Concrete models for the neutral current absorption scenario
and the current bounds can be found in
\cite{Dror+_2019a, Dror+_2019b, Dror+_2020, Ge+_2022,Ge:2023wye}.
There is still a significant allowed parameter space
for the neutral current fermionic DM absorption.

The neutral fermionic
absorption DM has a deep root in the history of
hunting neutrinos. After Pauli proposed neutrino
to explain the missing energy of beta decay \cite{Pauli:1991owm},
verifying its existence became a necessary and
urgent issue. Being neutral, it is very difficult
for neutrinos to be directly measured. Kan Chang Wang
suggested a two-body final state process with $K$-shell
electron capture, $e^- + {}^A_Z X \rightarrow \nu + {}^A_{Z-1} X$,
to probe neutrinos \cite{Wang:1942huv, Wang:1947}.
This is analogous to the
neutral current fermionic absorption
process with the electron mass converted to the nuclei
recoil energy for detection.
At that time, electrons and
nuclei were known particles while neutrinos were unknown.
Since neutrinos are well established now, the same process
can be used to probe the unknown fermionic DM particle
in the initial state.

The neutral fermionic DM absorption can occur with
either electron \cite{Dror+_2020, Ge+_2022}
or nucleus \cite{Dror+_2019a, Dror+_2019b, Li:2022kca}
as target. Since the absorption
process intrinsically consumes all the DM mass ($m_\chi$)
rather than just its kinetic energy as in the conventional
elastic scattering, the initial velocity is negligibly
small for halo DM. If the target particle is also
initially at rest, it then has a fixed recoil energy
$T_r = m^2_\chi / 2 (m_T + m_\chi)$ where $m_T$ is
the target mass \cite{Ge+_2022}.

However, the target electron is always bound in
an atom. Its binding energy is intrinsically at the
keV scale for heavy elements such as xenon. Being
comparable with the detection threshold,
it is then necessary to take the atomic effects
into consideration. The monoenergetic recoil energy
would broaden into a peak whose shape is determined
by the initial bound and final ionized electron
wave functions \cite{Dror+_2020, Ge+_2022}.
A consistent calculation framework of atomic
effects with second quantization
can be found in \cite{Ge:2021snv}.
Nevertheless, the initial nucleon motion and binding
effect have not been taken into consideration for the
nuclei target yet.

Since the initial DM particle is approximately at rest
with nonrelativistic velocity, the momentum transfer $\bm q$
is actually determined by the neutrino momentum $\bm p_\nu$.
With tiny mass, the neutrino momentum actually has the same
size as its energy. As a first estimation, let us assume
the initial nucleon is also at rest and leave the
effect of nucleon Fermi motion to be discussed later.
Then, the neutrino energy is the DM mass subtracting
the nucleon recoil energy,
\begin{align}
  |\bm{q}|
& = 
  |\bm{p}_\nu|
= 
  m_\chi
- \frac {m^2_\chi}{2 (m_N + m_\chi)}
\approx
  m_\chi,
\label{eq:3-momentum_transfer}
\end{align}
where $m_N \gg m_\chi$ is the target nucleon mass.
Note that the three-momentum transfer and its four-dimensional counterpart have
roughly the same value, 
$q^2 \equiv (p_\chi - p_\nu)^2 \approx - m^2_\chi$
where $p_\chi$ and $p_\nu$ are the DM and neutrino
momenta, respectively.
For $m_\chi \gtrsim 1 / R_N \sim \mathcal O(10)$\,MeV
that corresponds to the typical nuclei size $R_N \gtrsim 5$\,fm,
the fermionic DM absorption process can see individual nucleons.
For a typical DM direct detection experiment with
energy threshold
$T_r = m^2_\chi / 2 (m_A + m_\chi) \gtrsim 1$\,keV 
\cite{Ge+_2022} for a nuclei target with mass $m_A$,
the sensitive mass range is
$m_\chi \gtrsim \mathcal O(10)$\,MeV.
In other words, the individual nucleon can be readily
resolved and it is
necessary to consider the effect of incoherent scattering
with individual nucleons.

In this paper, we would explore
the incoherent neutral current fermionic DM absorption
with the Fermi motion of individual nucleon inside
the target nuclei.
In addition to kinematics, the effect on the
transition rate can be estimated with Pauli blocking
assuming Fermi gas description or form factor.

\vspace{1em}

\section{Fermi Gas and Recoil Energy Broadening}
With individual nucleons being resolved, the first effect
to be considered
is their Fermi motion inside nuclei and the broadening
of the corresponding recoil energy. We describe the incoherent
DM absorption using the impulse approximation, i.e.,
only a single nucleon participates in the process
\cite{Benhar+_2005}.
Furthermore, the nucleus is assumed to be a fully
degenerate Fermi gas with momentum distribution,
\begin{align}
    f({\bm{p}}_N)
=
    \frac{3}{4\pi p_F^3} \theta (p_F - |{\bm{p}}_N|),
\label{eq:RFG_momentum_distribution}
\end{align}
where $\theta (p)$ is a step function and
$p_F$ the Fermi momentum. With the prefactor
$3 / 4 \pi p^3_F$ that is the inverse of the
momentum space volume, $f(\bm p_N) d^3 \bm p_N = 1$.
The values of $p_F$ can be obtained from the electron-nucleus
scattering experiments \cite{Moniz+_1971, Bodek+_1981}.
A formula based on fitting these data can be
extracted from the GENIE code \cite{Andreopoulos+_2010}, 
\begin{align}
  p_F 
& = 
\left(
  0.27 
- \frac {1.13} A 
+ \frac {9.73}{A^2}
- \frac {39.53}{A^3}
\right)\mbox{GeV},
\label{eq:Fermi_momentum_GENIE}
\end{align}
where $A$ is the atomic mass. The proton and
neutron Fermi momenta $p_F^p$ and $p_F^n$ in
general differ from each other, especially
for heavy nuclei,
\begin{align}
  p_F^p
=
  p_F \sqrt[3]{\frac{2Z}{A}},
\quad
  p_F^n
=
  p_F \sqrt[3]{\frac{2(A-Z)}{A}},
\label{eq:Fermi_momentum_p_vs_n}
\end{align}
where $p_F$ is the Fermi momentum calculated
according to \geqn{eq:Fermi_momentum_GENIE}
and $Z$ is the charge number.
Once measured, the extent of broadening
can be used to identify whether the neutral
current fermionic DM absorption happens
with a proton or neutron target.

With a Fermi momentum $p_F \sim 250$\,MeV,
the typical nucleon velocity $v_N \sim 10^{-1}$
in the nucleus rest frame is much larger
than its DM counterpart $v_\chi \sim 10^{-3}$.
So assuming the DM particles to be at rest
in the lab frame is still a good approximation
while the initial nucleon needs to be relaxed
to have velocity $v_N$. Consequently, the nucleon
recoil energy $E_{N'} \equiv m_N + T_{N'}$ broadens to 
$E^+_{N'}\leq E_{N'} \leq E^-_{N'}$,
\begin{align}
\vspace{-2mm}
  E^\pm_{N'}
=
    \frac{1}{2s}
\left[ 
  (s + m_N^2) (E_N + m_\chi)
\pm
  (s-m_N^2) |\bm{p}_N|
\right],
\label{eq:recoil_energy_span}
\end{align}
with $E_N$ ($E_{N'}$) being the total energy
for the initial (final) nucleon $N$ ($N'$) and
$s \equiv m_N^2 + 2 m_\chi E_N +  m_\chi^2$.
Note that the energy boundaries $E^\pm_{N'}$
are independent of the nucleon momentum direction.
Inside the allowed energy range above, the
event rate has a flat distribution.

It is a good approximation to take $|\bm{p}_N| \ll m_N$.
Then for $m_\chi \ll m_N$, the kinetic energy boundaries
($T^-_{N'} < T_{N'} < T^+_{N'}$),
\begin{align}
  T^\pm_{N'}
&\approx 
  \frac {m^2_\chi + |\bm p_N|^2}{2 m_N}
\pm
  \frac {|{\bm p}_N|}{m_N} m_\chi
=
  \frac {(m_\chi \pm |\bm p_N|)^2}{2 m_N},
\label{eq:EN}
\end{align}
scale linearly with both the DM mass $m_\chi$
and the nucleon momentum $|\bm p_N|$. The energy
span, $2 m_\chi |\bm p_N| / m_N$, can reach
$\mathcal O(10)$\,MeV. Correspondingly, the
3-momentum transfer also broadens around
$|\bm{q}| \approx m_\chi$.

For the realistic measurement, the cross section $\sigma$ times
the initial relative velocity $v_N$ is the one that
directly enters event rate,
\begin{align}
  \langle \sigma v_N \rangle
=
  \int
  \frac{\overline{|\mathcal M|^2}}{4 m_\chi E_N}
  \frac {d T_{N'}}{8 \pi |\bm p_N|}
  f(\bm p_N) d^3 {\bm p_N},
\label{eq:vdsigma}
\end{align}
where $\overline{|\mathcal M|^2}$ is the spin
averaged matrix element. The nucleon velocity has
been canceled with the relative velocity in the
denominator of the cross section that originally
comes from the flux factor. Since the final-state
nucleon energy spans a range proportional to
$|\bm p_N|$ as shown above, the two-body phase
space integration $d \Phi_2 \equiv d T_{N'} / 8 \pi |\bm p_N|$
would not vanish with $|\bm p_N|$. Intuitively
thinking, the final-state phase space of an
absorption process is supported by the
consumed DM mass $m_\chi$,
$\int d \Phi = 2 m_\chi / m_N$,
rather than the initial nucleon momentum.

As an example, below we assume vector-type operator
$\mathcal{O}_V = (\bar{N} \gamma_\mu N)(\bar{\nu}_L \gamma^\mu\chi_L)$
for direct comparison with the experimental
analysis \cite{Dror+_2019b, Gu+_2022, Arnquist+_2022, Dai+_2022}.
The corresponding spin averaged matrix element is
\begin{align}
  \overline{|\mathcal M|^2}
& =
  \frac 4 {\Lambda^4}
  \left[ 
  (p_{N'}\cdot p_\nu) 
  (p_N\cdot p_\chi)
    \right.
\nonumber
\\  
&
\hspace{6mm}
+
  \left.
  (p_{N'}\cdot p_\chi) 
  (p_N \cdot p_\nu) 
  - m_N^2 
  (p_\nu \cdot p_\chi) 
  \right],
\end{align}
where $\Lambda$ is the cutoff scale for
the operator $\mathcal O_V$. When replacing
the momentum by variables in the lab frame,
$\overline{|\mathcal M|^2} \approx 4 m^2_\chi m^2_N (1 + 2 |\bm p_N|^2 / m^2_N) / \Lambda^4$.
With $|\bm p_N| / m_N \sim 0.25$, the
nucleon momentum can only induce $\lesssim 10\%$
correction. So it is a good approximation to take
$\overline{|\mathcal M|^2}
\approx 4 m^2_\chi m^2_N / \Lambda^4$
as a constant. Then the integration over $E_{N'}$
gives a constant
$\sigma v_N \approx m^2_\chi / 4 \pi \Lambda^4$
and hence the same for $\langle \sigma v_N \rangle$
\cite{Ge+_2022}.

\begin{figure}
\centering    
\includegraphics[width = \linewidth]{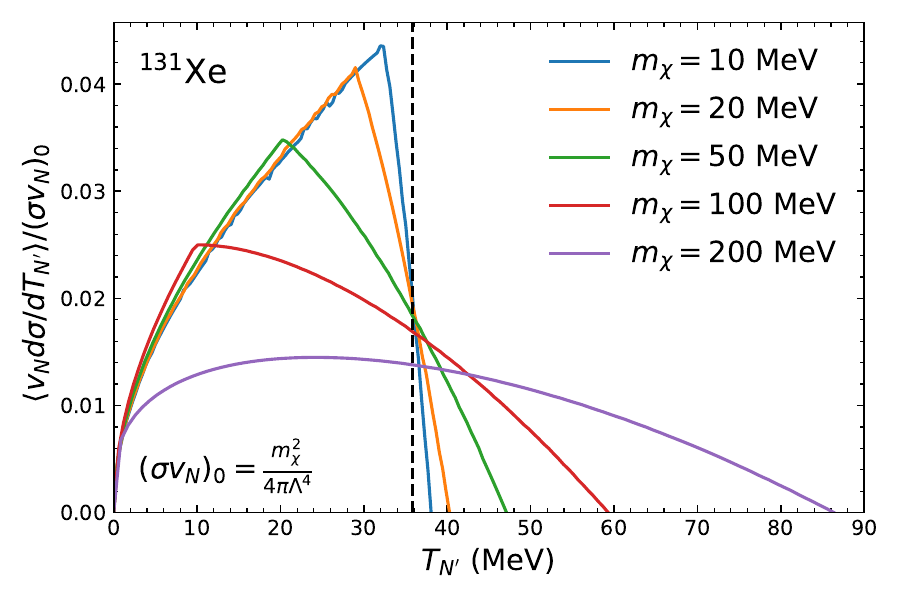}
\caption{The differential cross section of
the incoherent fermionic DM absorption with vector-type
effective operator $\mathcal{O}_V$ as a function
of the final nucleon kinetic energy $T_{N'}$.
With the Fermi momentum $p_F = 262$\,MeV
taken from the xenon nuclei $A = 131$, its corresponding
initial nucleon kinetic energy $T_N = 36$\,MeV
is shown with the dashed line.
}
\label{fig:dR_dT_smearing}
\end{figure}
%
Figure~\ref{fig:dR_dT_smearing} illustrates the effect of
Fermi motion on the differential cross section
$\langle d \sigma v_N / d T_{N'} \rangle$
for a xenon target ($A = 131$) with $p_F = 262$\,MeV.
For easy comparison, the differential cross section has
been normalized to the unit of
$(\sigma v_N)_0 \equiv m^2_\chi / 4 \pi \Lambda^4$.
With this normalization, the curves are independent of
the cutoff scale $\Lambda$ and the small DM mass region
is enhanced to make it more explicit.
As demonstrated in \geqn{eq:EN}, the span $m_\chi v_N$
of the nucleon recoil energy $T_{N'} \equiv E_{N'} - m_N
= T_N + m_\chi \pm m_\chi v_N$ where
$T_N \equiv E_N - m_N$ is the initial nucleon
kinetic energy,
scales linearly with the DM mass $m_\chi$. With
vanishing $m_\chi$, the recoil energy $T_{N'}$
reduces to a fixed value, $T_{N'} = T_N + m_\chi$.
For a given $T_{N'}$, the initial nucleon momentum
$|\bm p_N|$ is limited in the range,
$m_\chi - \sqrt{2 m_N T_{N'}} < |\bm p_N|
< m_\chi + \sqrt{2 m_N T_{N'}}$ by reversing
\geqn{eq:EN}. Then the averaged differential
cross section becomes
\begin{align}
  \left \langle \frac {v_N d \sigma}{d T_{N'}} \right \rangle
\approx
  \frac{3 \overline{|\mathcal M|^2}}{32 \pi m_\chi m_N}
  \frac {p^2_+ - p^2_-}{2 p^3_F}
\label{eq:diffXsec}
\end{align}
for $(\max \{ 0, m_\chi - p_F \})^2 / 2 m_N < T_{N'}
< (m_\chi + p_F)^2 / 2 m_N$. For a given $T_{N'}$,
the initial nucleon momentum limits are
$p_- \equiv |m_\chi - \sqrt{2 m_N T_{N'}}|$ and
$p_+ \equiv \min \{ m_\chi + \sqrt{2 m_N T_{N'}}, p_F\}$
which is always $m_\chi + \sqrt{2 m_N T_{N'}}$,
respectively. This approximation in \geqn{eq:diffXsec}
describes the behaviors in \gfig{fig:dR_dT_smearing}
quite well. With very light DM, the Fermi motion
significantly changes the monoenergetic recoil
to a wide spectrum with peak around the Fermi
energy $T_F \equiv p^2_F / 2 m_N$ shown as the
vertical dashed line. Heavier DM induces a wider
spectrum with a smooth peak and the DM recoil
energy can go beyond the Fermi energy by larger extent.

However, the nucleon bound inside nucleus cannot
be treated as a free particle. So the effect of
Fermi motion on scattering kinematics is not the
only consequence. By analogy with the atomic
effect for the fermionic DM absorption on
electrons inside atom \cite{Dror+_2020, Ge+_2022},
the cross section must also be affected. Below
we use two approaches, Pauli blocking and
form factors, to evaluate this effect.

\vspace{1em}

\section{Pauli Blocking}
Another consequence of the Fermi gas distribution in
\geqn{eq:RFG_momentum_distribution} is Pauli blocking.
Due to the Pauli exclusion principle, an initial-state
nucleon can only transit to the unoccupied final-state
levels with $|\bm{p}_{N'}| = |\bm{p}_{N} + \bm{q}| > p_F$.
Thus, the Pauli blocking effect can be taken into
account by introducing a step function
$\theta(|{\bf p}_{N'}| -  p_F)$ to \geqn{eq:vdsigma}.
The kinetic energy $T_F \equiv p^2_F / 2 m_N$
corresponding to the Fermi momentum has
been shown as the dashed vertical line in
\gfig{fig:dR_dT_smearing}.
Only those nucleons with kinetic energy
$T_{N'} \geq T_F$ can actually contribute to the process.

\begin{figure}[t]
\centering
\includegraphics[width = 1\linewidth]{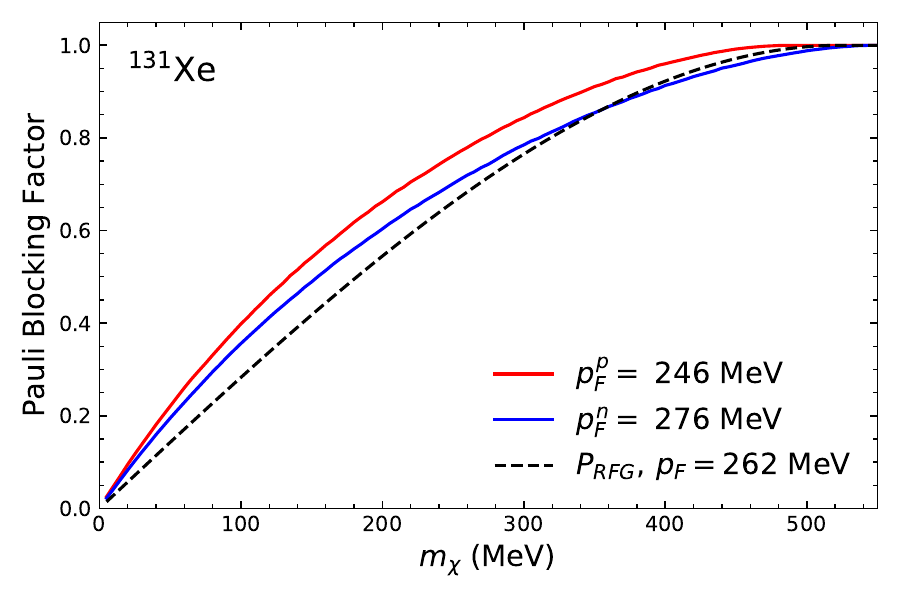}
\caption{The Pauli blocking factor on the averaged
cross section for protons (red) and neutrons (blue)
as functions of the DM mass $m_\chi$. For comparison,
the result with geometrical estimation is shown as
dashed line.
}
\label{fig:Pauli_blocking}
\end{figure}
Figure~\ref{fig:Pauli_blocking}
shows the Pauli blocking factor
defined as the ratio of the averaged cross sections with
and without the Pauli blocking effect. One may see that the
blocking effect is more significant for light DM whose mass
determines the energy gain. With larger $m_\chi$, it is
much easier for a nucleon to gain enough energy to
pass the Fermi motion threshold. With larger Fermi
momentum, neutrons have larger blocking effect than
the proton counterparts.

An intuitive estimation of the Pauli blocking effect
can also be obtained geometrically. For a
given momentum transfer $\bm q$, the fraction of nucleons
unaffected by Pauli blocking is determined by the overlap
between the initial Fermi sphere and the sphere shifted by
$\bm q$ \cite{Bodek_2021},
\begin{subequations}
\begin{align}
  P_{\rm RFG}
& =
  \frac{3}{4}\frac{|\bm q|}{p_F} - \frac{1}{16} 
  \left(
  \frac{|\bm q|}{p_F}
  \right)^3,
& \quad
  |\bm q| \leq 2p_F,
\\
  P_{\rm RFG}
&=
  1,
&\quad 
  |\bm q| > 2 p_F.
\label{eq:RFG_Pauli_suppression}
\end{align}
\end{subequations}
The corresponding Pauli blocking factor $P_{\rm RFG}$
for a relativistic Fermi gas (RFG) model is shown as
dashed line in \gfig{fig:Pauli_blocking}.
Since $|\bm{q}|\approx m_\chi$, and a typical Fermi momentum
is $p_F \approx 262$\,MeV for $^{131}$Xe, one can expect
suppression of DM absorption for $m_\chi \lesssim 500$\,MeV
which is roughly the place where
$P_{\rm RFG}$ starts to deviate from 1.
For $m_\chi = 100$\,MeV, only about $30\%$ of nucleons
will actually contribute. The geometrical picture gives
a quite good estimation of the Pauli blocking effect.

Nevertheless, the implementation of the Pauli exclusion principle
with a fully degenerate Fermi gas model according to
\geqn{eq:RFG_momentum_distribution} would lead to
an overestimation of the blocking effect. Naively thinking,
the final-state nucleon with momentum below $p_F$ should
not be completely forbidden since the nucleons inside
nuclei are not really fully degenerate. It is necessary
to find a different way, namely form factors for the
incoherent scattering, to get a better estimation.

\vspace{1em}

\section{Form Factors}
The coherent scattering occurs when the momentum
transfer is not large enough to resolve the individual
nucleons. So it is understandable that the coherent
scattering decreases while the incoherent counterpart
increases with momentum transfer. To be more specific,
the coherent scattering scales as $A^2 F^2(\bm q^2)$
and $F^2(\bm q^2)$ is
the coherent nuclear form factor while the incoherent one
scales as $A [1 - F^2(\bm q^2)]^2$ \cite{Bednyakov+_2018}.
The nuclear form factor is normalized such that $F(0) = 1$
(pointlike nucleus) and $F(\bm{q}^2)\to 0$
for $\bm{q}^2 \to \infty$ (fully incoherent case).

\begin{figure}
\centering
\includegraphics[width = 1\linewidth]{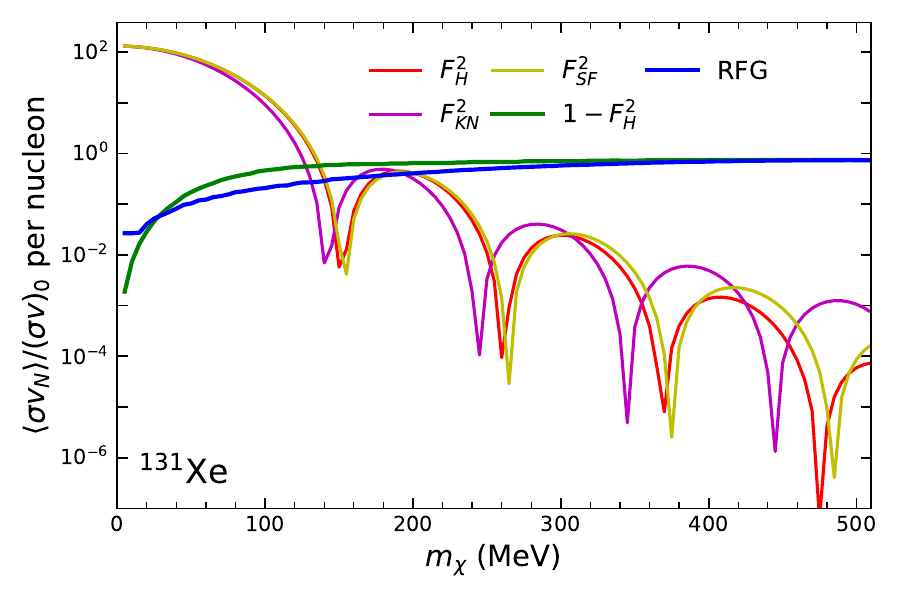}
\caption{The coherent ($F^2_{\rm H}$, $F^2_{\rm KN}$,
and $F^2_{\rm SF}$) and incoherent (green for $1 - F^2_{\rm H}$
and blue for Pauli blocking with RFG momentum distribution)
cross sections per nucleon on $^{131}$Xe target as
a function of the DM mass $m_\chi$.
}
\label{fig:RFG_vs_1-F2_A}
\end{figure}

Figure~\ref{fig:RFG_vs_1-F2_A}
illustrates the effect of
nuclear form factors on the coherent and incoherent
scattering cross sections for DM absorption on a $^{131}$Xe
target. To make fair comparison, the averaged cross section
per nucleon $\langle \sigma v_N \rangle / A$ is shown
in the unit of $(\sigma v_N)_0$. One may see that
the coherent scattering prevails for light DM with
$m_\chi \lesssim 100$\,MeV while the incoherent one
starts to dominate for heavier DM. Between the two
estimations of the incoherent contribution,
Pauli blocking with RFG momentum distribution (blue)
tends to give smaller values than the form factor
approach (green) due to the hard cut at the Fermi
momentum as mentioned above.

There are various parametrizations of the nuclear
form factors. We take the Helm, Klein-Nystrand,
and symmetrized Fermi-distribution schemes for comparison.
For completeness, we list their detailed parametrizations
below. The Helm form factor \cite{Helm_1956,Lewin+_1996},
\begin{align}
    F_{\rm H}(\bm{q}^2)
\equiv
    3\frac{j_1 (|\bm{q}|r_n)}{|\bm{q}| r_n} e^{-(|\bm{q}|s)^2 /2},
\end{align}
with $j_1 (x)$ being
the first order spherical Bessel function
contains a scale parameter,
$r_n \equiv \sqrt{c^2 + 7 \pi^2 a^2 / 3 - 5 s^2}$
where $c \equiv 1.23 A^{1/3} - 0.60$, $a = 0.52$,
and $s = 0.9$\,fm. For comparison, the Klein-Nystrand
form factor also has the $3 j_1(|\bm q| r_n) / |\bm q| r_n$
factor but the exponential suppression is replaced as
\cite{Klein+_1999}
\begin{align}
    F_{\rm KN}(\bm{q}^2)
\equiv
    3\frac{j_1 (|\bm{q}|r_n)}{|\bm{q}| r_n}
    \frac{1}{1 + |\bm{q}|^2 a_k^2}
\end{align}
with $a_k = 0.7$\,fm.
The symmetrized Fermi-distribution form factor has
a more complicated expression \cite{Piekarewicz+_2016},
\begin{align}
    F_{\rm SF}(\bm{q}^2)
& \equiv
    \frac{3}{|\bm{q}| c}
    \left[
    \frac{\sin (|\bm{q}|c)}{(|\bm{q}|c)^2}
    \frac{\pi |\bm{q}| a}{\tanh (\pi |\bm{q}| a)}
    -
    \frac{\cos (|\bm{q}|c)}
    {|\bm{q}|c)}
    \right]
\nonumber
\\
&\times
    \frac{\pi |\bm{q}| a}{\sinh (\pi |\bm{q}| a)}
    \frac{1}{1 + (\pi a /c)^2},
\end{align}
with $c \equiv 1.23 A^{1/3} - 0.60$\,fm and $a = 0.52$\,fm.

As shown in \gfig{fig:RFG_vs_1-F2_A},
the cross section of coherent scattering is quickly suppressed
by the momentum transfer that increases with the DM mass $m_\chi$.
For $m_\chi \lesssim 100$\,MeV, the difference among the adopted
form factors is not significant. In other words, the three
parametrizations work quite well in the coherent regime.
With larger momentum transfer, the difference can reach one
order of magnitude, especially around the dips. Since when reaching the
dip regions the incoherent scattering already starts to
dominate, these large differences would not have a significant
impact on the phenomenological studies in this paper.
For the incoherent scattering, the difference among form
factors is not visible since the corresponding form factor
is $1- F^2(\bm q^2)$ instead of $F^2(\bm q^2)$. Below
we would adopt the Helm form factor for further study.

\vspace{1em}

\section{Targets and Spin Dependence}
With amplification by a factor of $A$ for the coherent
scattering, the incoherent contribution starts to dominate only
for momentum transfer $|\bm q| \gtrsim 100$\,MeV.
This is
especially true for heavy nuclei such as xenon. But it is
not necessary for the coherent scattering always to be
amplified or even happen. There are two such cases.

\begin{figure}
\centering
\includegraphics[width=\linewidth]{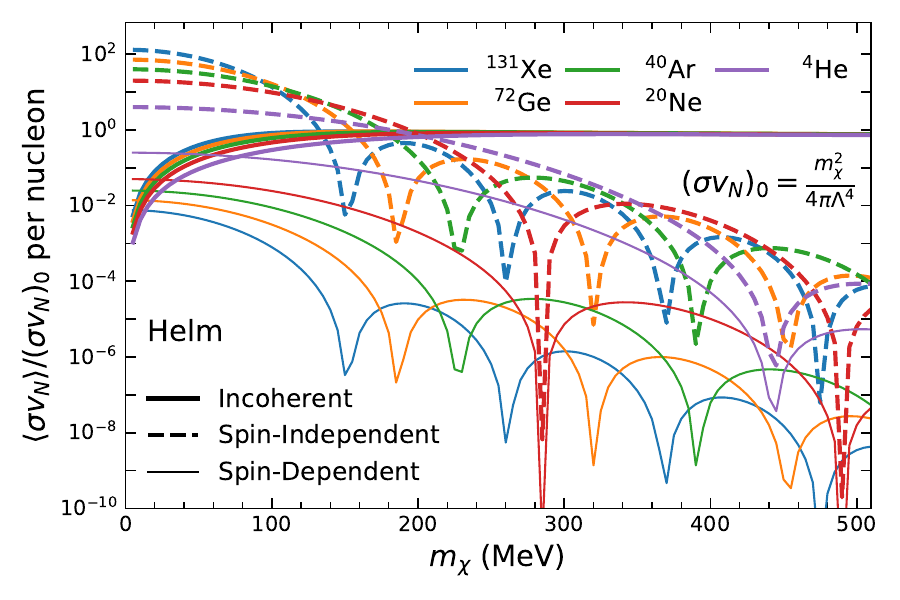}
\caption{The coherent (dashed and thin solid)
and incoherent (thick solid)
cross sections
per nucleon as functions of the DM mass
$m_\chi$ for
$^{131}$Xe (blue), $^{72}$Ge (orange), $^{40}$Ar (green),
$^{20}$Ne (dark red), and $^4$He (purple). For the coherent
ones, both spin-independent (dashed)
and spin-dependent (thin solid) cases have been shown.
}
\label{fig:tot_event_rates_var_targets_FF}
\end{figure}

One way is using a light nuclei that has smaller nucleon
number $A$. 
Figure \ref{fig:tot_event_rates_var_targets_FF}
shows
the coherent cross section per nucleon for the typical
nuclei (including $^{131}$Xe, $^{72}$Ge, $^{40}$Ar,
$^{20}$Ne, and $^4$He) used for DM direct detection
experiments as dashed lines. With smaller nucleon number,
the amplification of the coherent scattering compared with
its incoherent counterpart would not be that large. The
coherent cross section on the left-hand side keeps
decreasing with smaller nuclei. For heavier DM mass,
the form factor starts to have a dip around
$m_\chi \sim 4.5 / r_n$. Since the radius parameter
$r_n$ increases with the atomic mass $A$, a heavier
nucleus has its first dip at smaller $m_\chi$.
Consequently, the dominance of the incoherent scattering
appears at lighter DM mass. The heavier nucleus actually
has the advantage of seeing the incoherent scattering for
$m_\chi \gtrsim 100$\,MeV.

Another way is the spin dependence of the effective
DM absorption operators. The coherent enhancement $A^2$
happens for the spin-independent nucleon vertex. Being
independent of the nucleon spin, the individual scattering
matrix elements add up coherently before being squared
to give the cross section. However, the amplification
is lost for a spin-dependent nucleon vertex. For comparison
with the vector operator
$\mathcal{O}_V = (\bar{N} \gamma_\mu N)(\bar{\nu}_L \gamma^\mu\chi_L)$ used in our earlier discussions,
its axial vector counterpart
$\mathcal{O}_A = (\bar{N} \gamma_\mu \gamma_5 N)(\bar{\nu}_L \gamma^\mu \gamma_5 \chi_L)$
would lead to a spin dependent nonrelativistic
nucleon vertex. The dashed and dotted lines in
\gfig{fig:tot_event_rates_var_targets_FF} shows the
spin-independent $\mathcal O_V$ and spin-dependent
$\mathcal O_A$ contributions. Without amplification,
the cross section on the left-hand side significantly
reduces and its value per nucleon becomes inversely
proportional to the atomic mass $A$.
The coherent vs incoherent equality happens at much
smaller DM mass now. It is still true that
for a heavier nucleus the incoherent
scattering starts to dominate for smaller DM mass.

\begin{figure}[t]
\centering
\includegraphics[width=\linewidth]{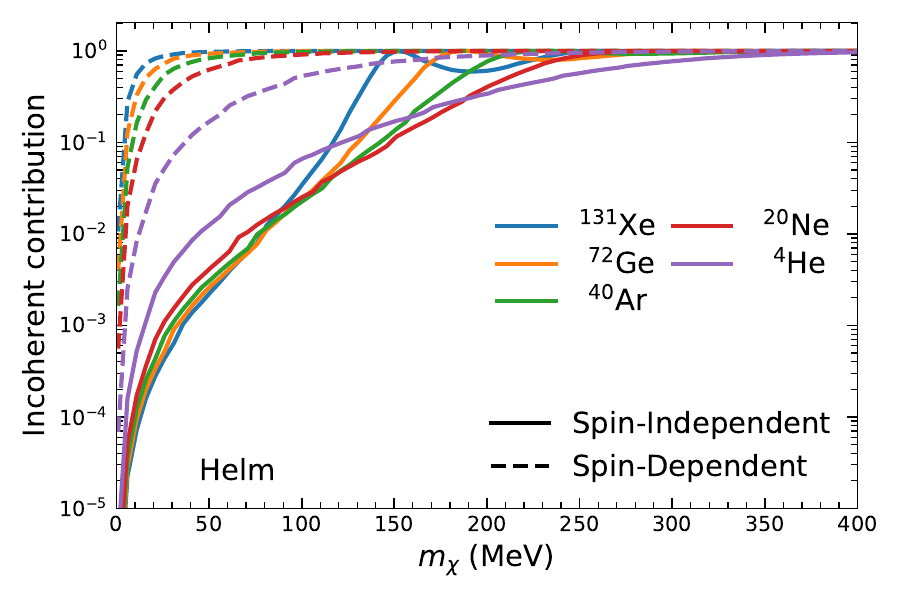}
\caption{The relative contribution of the
incoherent fermionic DM absorption for different
nuclear targets including
$^{131}$Xe (blue), $^{72}$Ge (orange), $^{40}$Ar (green),
$^{20}$Ne (dark red), and $^4$He (purple).
For the coherent part, the spin-independent
vector operator $\mathcal O_V$ with Helm form
factor is considered for comparison.
}
\label{fig:rel_contribution_var_targets}
\end{figure}
Figure \ref{fig:rel_contribution_var_targets}
shows the
relative contributions of the incoherent scattering
as functions of the DM mass $m_\chi$. For comparison,
the coherent scattering with spin-independent
vector operator $\mathcal O_V$ is shown as solid
lines while
for the spin-dependent axial vector operator $\mathcal O_A$
the dashed curves significantly move up.
One may see that the incoherent contribution increases
with the DM mass $m_\chi$ which roughly corresponds
to the momentum transfer. The small wiggle when
approaching 1, especially for the $^{131}$Xe target
that is shown as blue line, is caused by the dip and
peaks in the Helm form factor. For all cases, the
incoherent scattering becomes important for
$m_\chi \gtrsim 100$\,MeV with spin-independent operator.
Figure \ref{fig:rel_contribution_var_targets}
also shows the results with spin-dependent operator as
dashed lines. The fraction of incoherent scattering
is then significantly enhanced for light DM mass
$m_\chi \lesssim 100$\,MeV as expected from
\gfig{fig:tot_event_rates_var_targets_FF}.

\begin{figure}[t]
\centering
\includegraphics[width=\linewidth]{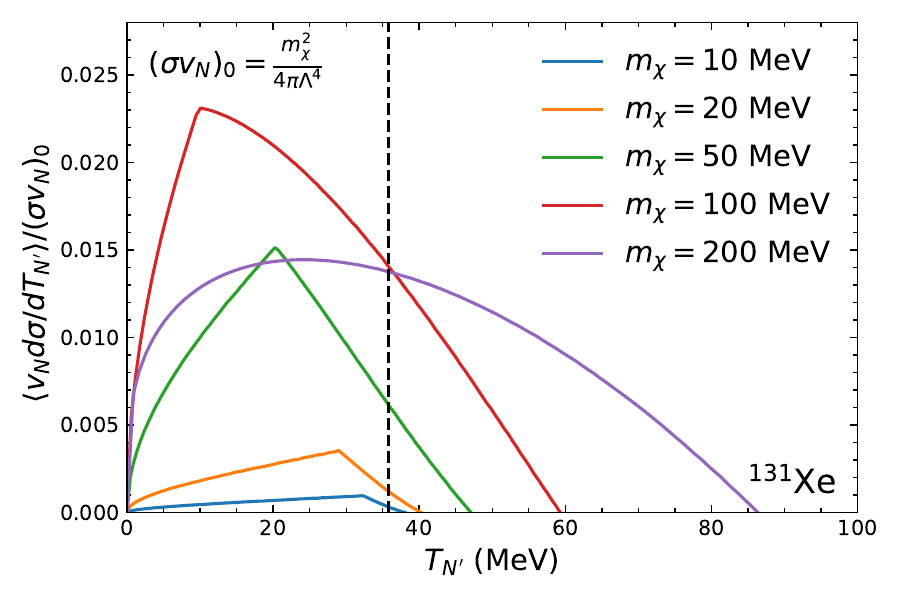}
\caption{The differential cross section for the
incoherent absorption process in the unit of
$(\sigma v_N)_0 = m^2_\chi / 4 \pi \Lambda^4$. }
\label{fig:diff_ev_rates_var_targets_PB}
\end{figure}
It is then necessary to study the differential cross
section of the incoherent scattering with the Helm form factors
as shown in \gfig{fig:diff_ev_rates_var_targets_PB}.
Comparing it with \gfig{fig:dR_dT_smearing}, the cross
section for light DM is significantly suppressed by
the incoherent form factor $1 - F^2(\bm q)$. Note that
the curves shown in both 
Figs. \ref{fig:dR_dT_smearing} and \ref{fig:diff_ev_rates_var_targets_PB}
are in the unit
of $(\sigma v_N)_0 \equiv m^2_\chi / 4 \pi \Lambda^4$
that is already suppressed by the light DM mass.
These features are consistent with the quick decreasing
of the relative contribution of the incoherent scattering
on the left side of \gfig{fig:rel_contribution_var_targets}.

\vspace{1em}
\section{Projected Sensitivities}
\begin{figure}[t]
\centering
\includegraphics[width=\linewidth]{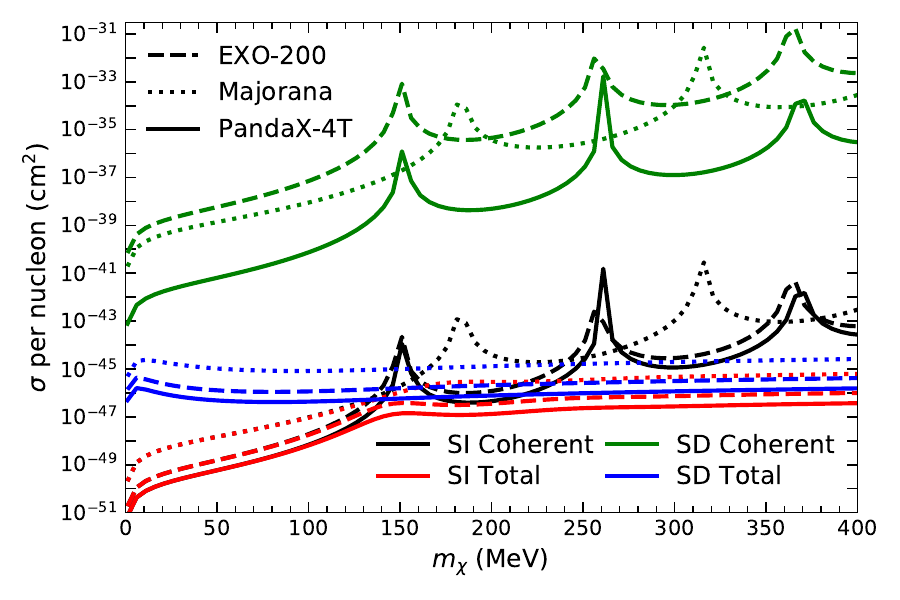}
\caption{The projected exclusion curves for the current experiments EXO-200 \cite{Al_Kharusi+_2023} (dashed), Majorana Demonstrator \cite{Arnquist+_2022} (dotted), and PandaX-4T \cite{Gu+_2022} (solid). For each experiment,
the four projected sensitivities are obtained for
the coherent-only (coherent) or total (total)
spin-independent (SI) or spin-dependent (SD)
interactions.}  
\label{fig:exclusion_plot}
\end{figure}

Finally, we demonstrate how the experimental sensitivities change if the incoherent channel is taken into account. 
As concrete examples, we consider the current
experiments EXO-200 \cite{Al_Kharusi+_2023} (exposure $234.1$\,kg$\times$yr), Majorana Demonstrator \cite{Arnquist+_2022} ($37.5$\,kg$\times$yr), and PandaX-4T \cite{Gu+_2022} ($0.63$\,t$\times$yr). 
Figure \ref{fig:exclusion_plot}
shows the estimated upper limits on the cross section per nucleon, assuming no more than $10$ events occur in the experiment. 

As discussed above, for the spin-independent interaction, the coherent and incoherent contributions become comparable at $m_\chi \sim 150$\,MeV. Accordingly, for $m_\chi \gtrsim 150$\,MeV, the limits obtained with both contributions (red) start to be stronger than those obtained for the coherent-only case (black).

For the spin-dependent interactions, the difference between
the both-channels (blue) and coherent-only (green) scenarios is even more dramatic. The incoherent contribution already starts to dominate at $m_\chi \sim 30$\,MeV. This is caused by two reasons. First, some isotopes such as ${}^{136}$Xe with even-even nuclei
cannot contribute to the coherent scattering while
all isotopes can contribute to the incoherent one.
Second, even for those nuclei that can contribute,
the coherent channel does not scale with the atomic
mass $A$ while its incoherent counterpart scales linearly
with $A$. The difference is especially pronounced
for EXO-200 and Majorana which are
enriched in even-even isotopes.

Therefore, when the incoherent absorption is taken
into account, the sensitivities to both the
spin-independent and spin-dependent interactions improve.
The improvement is especially significant for the
spin-dependent case.

\vspace{1em}

\section{Discussion and Conclusion}
For the fermionic absorption process, not just the mass
conversion to the final-state kinetic energy is efficient
but also for the momentum transfer. Being roughly the same
size as the absorbed DM mass, the momentum transfer
can already enter the incoherent regime for
a DM mass $m_\chi \gtrsim 100$\,MeV with a spin-independent
operator. For comparison, the classical direct detection
channel with elastic scattering can only provide a momentum
transfer $|\bm q| \sim m_\chi v_\chi$ that is
suppressed by the DM velocity $v_\chi \sim 10^{-3}$ that
is just 0.1\% of the speed of light.
This feature should apply for not just the
fermionic absorption process when the DM mass is fully
converted but also any exothermic DM scenario.

Our paper is a first attempt to demonstrate and estimate the
effects of incoherent scattering. Although a heavier
nucleus tends to have larger scattering
cross section in the coherent regime with a
spin-independent operator, it has the
advantage to see the coherent-incoherent equality
at a smaller DM mass. For a spin-dependent operator,
the coherent scattering cross section per nucleon
actually decreases with the nuclei mass number and
it is easier for a heavier nuclei target to observe
the incoherent scattering.

The Pauli blocking
treatment with relativistic Fermi gas model is quite
rough but gives a very good illustration. In comparison,
the form factor approach gives more features and is a
commonly used way in the study of DM direct detection.
Future improvement would be a realistic nuclear calculation
of the transitions caused by the fermionic DM absorption
which would be carried out in our next study.

\begin{acknowledgments}
The authors would like to thank Yong Du and Yakun Wang for useful discussions.
The authors are supported by the National Natural Science
Foundation of China (Grants No. 12375101, No. 12425506, No. 12090060, and No. 12090064) and
the SJTU Double First Class start-up fund (Grant No. WF220442604).
S.F.G. is also supported as an affiliate member of Kavli IPMU, University of Tokyo.
\end{acknowledgments}

\end{document}